\documentclass[aps,preprint,showkeys]{revtex4}
\usepackage{amsfonts}
\usepackage{amsmath}
\usepackage{amssymb}
\usepackage{graphicx}
\usepackage{subfigure}
\usepackage{txfonts}
\usepackage{makeidx}

\begin{document}

\title{\textbf{Halogenation of SiC for band-gap engineering and excitonic
functionalization}}
\author{L. B. Drissi$^{1,2}$, F. Z. Ramadan$^{1}$, S. Lounis$^{3,\ast}$}
\affiliation{{$^1$ LPHE, Modeling \& Simulations, Faculty of Science, Mohammed V
University in Rabat, Morocco}}
\affiliation{{$^2$ CPM, Centre of Physics and Mathematics, Faculty of Science,
Mohammed V University in Rabat, Morocco}}
\affiliation{{$^3$ Peter Gr\"{u}nberg Institut and Institute for Advanced Simulation,
Forschungszentrum J\"{u}lich and JARA, D-52425 J\"{u}lich, Germany}. }

\begin{abstract}
The optical excitation spectra and excitonic resonances are investigated in
systematically functionalized SiC with Fluorine and/or Chlorine utilizing
density functional theory in combination with many-body perturbation theory.
The latter is required for a realistic description of the energy band-gaps
as well as for the theoretical realization of excitons. Structural,
electronic and optical properties are scrutinized and show the high
stability of the predicted two-dimensional materials. Their realization in
laboratory is thus possible. Huge band-gaps of the order of 4 eV are found
in the so-called GW approximation, with the occurrence of bright excitons,
optically active in the four investigated materials. Their binding energies
vary from 0.9 eV to 1.75 eV depending on the decoration choice and in one
case, a dark exciton is foreseen to exist in the fully chlorinated SiC. The
wide variety of opto-electronic properties suggest halogenated SiC as
interesting materials with potential not only for solar cell applications,
anti-reflection coatings or high-reflective systems but also for a possible
realization of excitonic Bose-Einstein condensation. \newline
\newline
\end{abstract}

\maketitle

\section{Introduction}

Graphene has been largely studied and developed as a promising candidate for
both fundamental research and advances in nanotechnology. This carbon
nanosheet carries many unique properties over conventional materials such as
high electrical conductivity, high electron mobility at room temperature 
\cite{novoselov, geim}, and excellent optical transmittance \cite{trans} in
the visible light range, which render this material highly promising for
transparent conducting electrodes \cite{nature} and solar cells \cite{abs}.
Moreover, transparent electrode based-graphene yields white organic
light-emitting diodes with brightness and efficiency sufficient for general
lighting \cite{han}.

Functionalization of graphene in order to tune its properties by different
combinations of several types of materials is a large stream of activities
in what we would name graphinology. For instance, oxygenation\cite{OC} and
hydrogenation \cite{ane} has inspired intense search for novel
graphene-based materials. Fluorographene and chlorographene, engineered
under ambient conditions \cite{FC,clC}, are highly stable thanks to the
large electronegativity of fluorine and chlorine adsorbates \cite%
{halogengraphene}. Interestingly, in graphene halides, the binding of F- and
Cl-atoms to the graphene surface leads to covalent bonds that change the
hybridization state from $sp^{2}$ to $sp^{3}$ \cite{halogen2}. However,
carbon orbitals retain their $sp^{2}$ hybridization during the adsorption of
bromine and iodine. A crucial impact of halogens is in significantly
decreasing the gap energy of their hydrogenated counterparts, making the new
derivatives suitable for channel materials \cite{halogengraphene}.
Interestingly, diatomic halogen molecules (I$_{2}$, Br$_{2}$) change carrier
concentration in graphene without significant reduction of their mobility
near the Dirac point \cite{IBr}. Fully co-decorated graphene sheets with
chemical species such as HF and HCl were also reported \cite{HFCL}. The
resulting structures present a low level of disorder, high stability and
band-gaps of about $3$eV. Many of these properties are shared with 2D
counterparts of graphene, such as silicene halides (see e.g. \cite%
{vogt,houssa,ClSii,silicenehal}). Interestingly, effective carrier masses in
the latter complex are comparable in certain directions to those of silicon
and show a relatively high mobility \cite{zan}.

Our investigation is related to a relatively new 2D material, SiC, based on
graphene and silicene, so-called silicene-graphene, which has been recently
synthesized \cite{Sicexp}, following theoretical predictions \cite{oor}-\cite%
{Shi}. SiC offers an alternative to overcome limitations associated with
graphene electronic technologies. For example, it has a non-zero direct band
gap energy with a high exciton binding energy $E_{b}$ of $0.81eV$ as
predicted by theory using GW built from localized density approximation- 
\cite{drissi2}. The gap opening originates from the alternated arrangement
of carbon and silicon in a two-dimensional honeycomb lattice which breaks
the sublattice symmetry (see e.g. \cite{drissi1}). Thus, SiC hybrid could
also be a remarkable candidate for novel type of light-emitting diodes as it
shows improved photo-luminescence compared to its sphalerite or wurtzite
counterparts \cite{light}.

Our interest is focused further on the ability to engineer excitons, which
in nanostructures can have large binding energies. These excited
quasi-particles (QP) made of an electron bound to a hole are pivotal in
optoelectronic and thus, in photovoltaics, photoluminescence (see e.g. \cite%
{scholes}) and potentially even in quantum information technology \cite{poem}%
. Excitons can be of the bright-type, i.e. optically active and thus of
strong interest, or of the dark-type, i.e. optically inactive. The
difference between the two types of excitons hinges on the spin-alignment of
the electron and of the hole. However as highlighted by Poem et al.\cite%
{poem}, dark excitons could also be of use in opto-applications directed
towards quantum information technology.

Contrary to graphene and silicene, less attention has been devoted to
functionalization of 2D-SiC with atomic decoration. Interestingly, full
hydrogenation of SiC, where H-atoms passivate the surface via the unpaired
electrons of the substrate atoms, yields to a buckled system with an
insulating character \cite{drissi1}. The new material was predicted to be
mechanically stable with strong resistance to the in-plane strains and shear
waves can propagate faster than graphene along a specific direction \cite%
{drikaw1}. Partial hydrogenation, however, has the opposite effect on the
electronic structure since the gap is reduced in favor of ferromagnetism 
\cite{drissi1}. Similar to hydrogenation, complete fluorination tailors
electronic properties of SiC and semi-fluorination has the additional
advantage to induce novel magnetoelectric properties \cite{drissi3}.

The goal of our work is to provide a first-principles based investigation of
opto-electronic properties of 2D-SiC functionalized through halogenation.
The latter allows tuning the band gap together with a generation of excitons
having a high binding energy. Atomic decoration is made with either F, Cl or
a combination of both elements. For an accurate quantitative estimation of
gap energies of 2D materials required for practical applications, it is
essential to utilize many-body perturbation theory within GW approximation
that goes beyond the usual approximations LDA and GGA \cite{LOUI,EE}.
Many-body effects, such as e-e correlations, included in the GW
approximation correct the optical response, while e--h interactions as
described in the framework of the Bethe-Salpeter equation (BSE) are required
for the excitonic effects \cite{bse}.

We consider four configurations: (i) full fluorinated silicene/graphene
hybrid (F-SiC-F), (ii) full chlorinated hybrid (Cl-SiC-Cl), (iii) mixed
Cl-SiC-F where F-atom decorate the carbon atoms while Cl-atom the silicon
atoms and (iv) the opposite mixed configuration F-SiC-Cl. Phonon frequencies
analysis reveals the stability of all conformers. Our calculations
demonstrate that halogen-atoms can greatly enhance excitonic effects in
silicene/graphene hybrid, where the binding energies of the four
configurations are higher than of pristine. The highest exciton binding
energy of 1.75eV is found in F-SiC-Cl. We recall that bound excitons in
fluorographene have a large formation energy, $1.96$eV, of bound excitons 
\cite{hh} compared to $1.25$eV obtained for chlorographene \cite%
{chlorographene}, while in fluorinated silicene, the binding energy is of $%
1.48$eV \cite{fluorosilicene}. The reflectivity spectra demonstrate that
these materials are transparent in the visible region.

\section{Computational details}

To study phonon spectra, structural, electronic and optical properties of
halogenated SiC, we use the Quantum espresso (QE) simulation package \cite%
{A8} based on DFT and employing the GGA functional of Perdew-Burke-Ernzerhof
(PBE) \cite{PBE}. A norm-conserving pseudo-potential description \cite{123}
is used. A kinetic energy cutoff of 65Ry was applied\textrm{\ }for the
plane-wave basis.\textrm{\ }In the Monkhorst-Pack grid, the Brillouin-zone
integration was carried out at $16\times \ 16\times \ 1$ k-points for
describing the band structure while the stabilities of structures are
examined by calculating binding and formation energies and performing phonon
dispersions using $20\times 20\times 1$ k-points. The optimized unit cell is
obtained by minimizing the total energy as a function of the lattice
parameter. All the structures are relaxed using a criterion of forces and
stresses on atoms; until the energy change is smaller less than $10^{-4}$eV.

The energy band gap is corrected using non-self consistent GW calculations
implemented in YAMBO program suite \cite{YAMBO}. In the GW approximation,
the exchange-correlation potential $V_{xc}$ used in the DFT is replaced by a
nonlocal, energy-dependent self-energy operator. The first order
quasiparticle (QP) corrections are obtained using Hedin \cite{hedi} GW
approximation, while excitonic effects are treated by solving the
Bethe-Salpeter equation (BSE) \cite{bse}. In our calculations of QP
energies, we use a $12\times 12\times 1$ k-point mesh. The same mesh is used
to evaluate the dielectric function in both the random phase approximation
(RPA) \cite{RPA} and in the Tamm-Dancoff approximation \cite{I} that
considers only the resonant part of the BSE.

The charges of the ions were calculated according to Bader formalism that
gives the charge of each atom constituting the material by integrating over
the gradient paths of its electronic density \cite{Bader}.

\section{Results and discussion}

\begin{figure}[tbph]
\begin{center}
\hspace{0cm}\includegraphics[width=17cm]{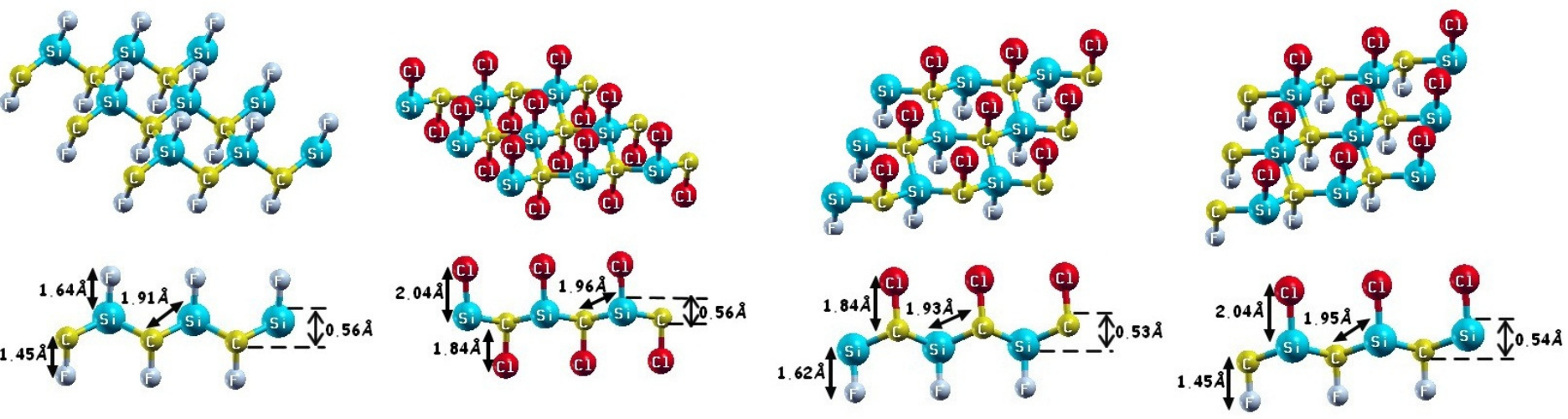}
\end{center}
\caption{{\protect \small Top and side views of the fully relaxed structures
showing interatomic distances.}}
\label{fig1}
\end{figure}

Silicene-graphene hybrid is stable in a planar hexagonal lattice with all
bond angles $\mathrm{Si}\widehat{\mathrm{C}}\mathrm{Si}$ and $\mathrm{C}%
\widehat{\mathrm{Si}}\mathrm{C}$\ equal to $120^{\circ }.$ The bond length
C-Si of $1.78\mathring{A}$ is bounded below by the one of graphene and above
by the one of silicene \cite{oor,drissi1}$.$ In the following, we study full
halogenated SiC where all the species adopt a chair configuration because it
is the most stable one with respect to zigzag, boat and armchairs \cite%
{silicenehal,chlorographene}.

Four structures in chair configuration, namely F-SiC-F, F-SiC-Cl, Cl-SiC-F
and Cl-SiC-Cl, are considered. As shown in Fig.\ref{fig1}, full fluorinated
(chlorinated) SiC referred as F-SiC-F (Cl-SiC-Cl), are obtained by attaching
F (Cl) atoms to C atoms in one side and to Si atoms in the opposite side of
the plane. In chlorosilicene-fluorographene hybrid, Cl-SiC-F, all carbon
atoms are decorated by fluorine atoms and all Si atoms bond chlorine atoms
forming 1up/1down fashion on either side of the sheet. In F-SiC-Cl, the role
of coadsorbed F- and Cl-atoms are inverted.

\begin{table}[tbp]
\begin{tabular}{l||llll|cc|ll|lll}
\hline
Formers & d$_{\mathrm{Si}-\mathrm{C}}$ & d$_{\mathrm{Si}-\mathrm{ad}}$ & d$_{%
\mathrm{C}-\mathrm{ad}}$ & $\Delta $ & $E_{B}$ & {$E_{F}$} & $E_{\mathrm{gap}%
}^{\mathrm{GGA}}$ & $E_{\mathrm{gap}}^{\mathrm{GW}}$ & $E_{b}^{e-h}$ & $M_{%
\mathrm{eff}}$ & $r_{Bohr}$ \\ \hline
Pure & \multicolumn{1}{||c}{$1.78$} & \multicolumn{1}{c}{-} & 
\multicolumn{1}{c}{-} & $0.00$ & - & - & $2.52$ & $3.53$ & $1.05$ & $0.53$ & 
$2.14$ \\ \hline
F-SiC-F & $1.91$ & $1.64$ & $1.45$ & $0.56$ & $-2.32$ & $-1.45$ & ${1.87}$ & 
${4.47}$ & $1.74$ & $0.40$ & $2.34$ \\ 
F-SiC-Cl & $1.93$ & $1.62$ & $1.84$ & $0.53$ & $-1.93$ & $-1.05$ & ${1.89}$
& ${4.71}$ & $1.75$ & $0.40$ & $2.24$ \\ 
Cl-SiC-F & $1.95$ & $2.04$ & $1.45$ & $0.54$ & $-1.71$ & $-0.83$ & ${2.05}$
& ${4.26}$ & $0.9$ & $0.22$ & $4.40$ \\ 
Cl-SiC-Cl & $1.96$ & $2.04$ & $1.84$ & $0.57$ & $-1.33$ & $-0.44$ & ${2.25}$
& ${4.38}$ & $1.31$ & $0.26$ & $3.36$ \\ \hline
\end{tabular}%
\caption{{\protect \small Structural parameters and different energies
characterizing the pure and the four types of decorated SiC. The interatomic
distances d, the buckling parameter }$\Delta ${\protect \small \ and Bohr
radius, }$r_{Bohr}${\protect \small , are given in }$\mathring{A}$%
{\protect \small . The Binding energy }$E_{B}${\protect \small , formation
energy }$E_{F}$,{\protect \small \ gap energy }$E_{gap}${\protect \small \ and
excitonic binding }$E_{b}^{e-h}${\protect \small \ energies are all in eV
while the effective mass of the exciton, }$M_{\mathrm{eff}}${\protect \small %
, is given in terms of the bare electron mass }$m_{0}$.}
\end{table}

Structure parameters, listed in Table 1, reveal that both C- and Si-atoms
are displaced along the normal direction from the plane of SiC layer as they
are fourfold coordinated and form sp$^{3}$ hybridization with their
neighbors. As a consequence, the halogenated structures are buckled and
their bonds Si--C are stretched, ranging in between $7\%$ and $10\%$, with
respect to pure SiC hybrid. The interatomic distances $d_{\mathrm{F}-\mathrm{%
C}}/d_{\mathrm{Cl}-\mathrm{C}}$ and $d_{\mathrm{F}-\mathrm{Si}}/d_{\mathrm{Cl%
}-\mathrm{Si}},$ that increase with the increase of the atomic number of
halogen elements, are in line with data calculated for halogenated graphene 
\cite{halogengraphene} and halogenated silicene \cite{silicenehal}.

Relative stability of configurations is studied first by evaluating their
binding $E_{B}$ and formation $E_{F}$ energies with respect to the energies
of pure SiC and of the isolated molecules made of the adsorbate. The
corresponding energies are expressed as \cite{energy}:%
\begin{equation}
E_{B}=\frac{1}{N}\left[ E_{T}-E_{\mathrm{SiC}}-n_{\mathrm{ad}}E_{\mathrm{ad}}%
\right] \text{ \  \  \ and \  \ }E_{F}=\frac{1}{N}\left[ E_{T}-E_{\mathrm{SiC}}-%
\frac{1}{2}\left( n_{\mathrm{ad}}E_{\mathrm{ad}_{2}}\right) \right]
\end{equation}

where $N$ is the total number of atoms per unit cell, $E_{T}$ is the total
energy of the (co)-decorated structure and $E_{\mathrm{SiC}}$ is the energy
of pristine SiC, n$_{\mathrm{ad}}$ is the number of adsorbated atoms in the
supercell under consideration, $E_{\mathrm{ad}}$ and $E_{\mathrm{ad}_{2}}$
describe respectively the energy of an isolated adsorbed atom and the energy
of the corresponding molecule.

Data in Table 1 shows that all configurations have negative formation and
binding energies. Therefore, the structures are stable and the adsorptions
are exothermic processes that could then be synthesized in experiments.
Fluorine adsorbed on SiC sheet generates the most stable structure among
halogen adsorbates. This result is in good agreement with previous works on
halogenated graphene and halogenated silicene \cite%
{silicenehal,chlorographene}.

\begin{figure}[tbph]
\begin{center}
\hspace{0cm} \includegraphics[width=17cm]{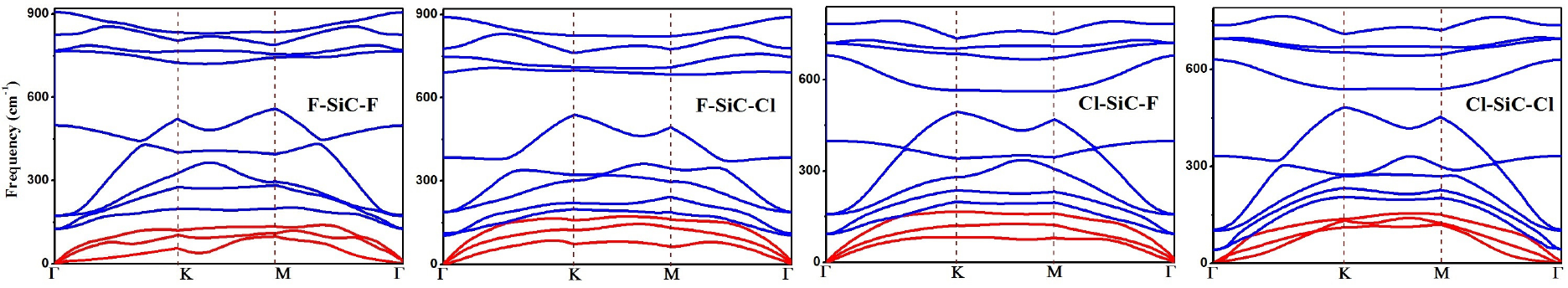}
\end{center}
\par
\vspace{-0.5cm}
\caption{{\protect \small Frequencies of the phonon modes of halogenated SiC
hybrids. The optical branches are shown in red while the acoustical ones are
in blue.}}
\label{2}
\end{figure}

Fig.\ref{2} displays dispersions of phonon modes, which is a reliable test
for the examination of thermodynamic stability of halogenated SiC hybrids.
The analysis of the curves shows that all phonon branches do not have
imaginary frequency along any high-symmetry direction of the Brillouin zone.
This is a signature of stability of the four structures. Each phonon
spectrum includes $12$ phonon bands, $3$ acoustic and $9$ optical. Phonon
frequencies soften monotonically from full fluorinated to full chlorinated
SiC due to the increasing atomic weight. It is observed that acoustic and
optical branches are separated by a band gap in F-SiC-F. As a result, it is
rather difficult to satisfy the energy conservation law during the
phonon-phonon scattering between acoustic and optical modes in this
structure.

\begin{table}[tbp]
\begin{tabular}{l||l|l|l|l}
\hline
& F-SiC-F & Cl-SiC-Cl & F-SiC-Cl & Cl-SiC-F \\ \hline
C & $-0.96$ & $-0.96$ & $-0.96$ & $-0.97$ \\ 
Si & $+3.86$ & $+3.85$ & $+3.85$ & $+3.88$ \\ 
ad$_{/\mathrm{C}}$ & $-0.69$ & $-0.20$ & $-0.14$ & $-0.72$ \\ 
ad$_{/\mathrm{Si}}$ & $-0.98$ & $-0.97$ & $-0.97$ & $-0.97$ \\ \hline
\end{tabular}%
\caption{{\protect \small Charge transfer involved in the four studied
compounds: (+) sign denotes loss of electrons in opposite to (-) sign.}}
\end{table}

To grasp the bonding nature among atoms, Table 2 summarizes values of charge
transfer for the four species calculated using Bader charge analysis.
C-atoms are charge acceptors with $\sim 0.96e$ obtained from their
Si-surroundings. It follows that the bond between Si and C is ionic.
Moreover, the adsorbates gain an important amount of charge from their
Si-atoms neighbor compared to the smaller electrons density transferred from
C-atoms. This behavior can be attributed to Pauling electronegativities of F
(3.98), Cl (3.16), C (2.55) and Si (1.9), since electrons are pulled away
from atoms having lower electronegativity towards atoms with a higher one.
Except C-Cl that forms a covalent bond, all the other bonds in the four
configurations are ionic, which makes them suitable for many applications,
because ionic materials are strong, hard and also they have a high melting
point.

\begin{figure}[tbph]
\begin{center}
\hspace{0cm} \includegraphics[width=17cm]{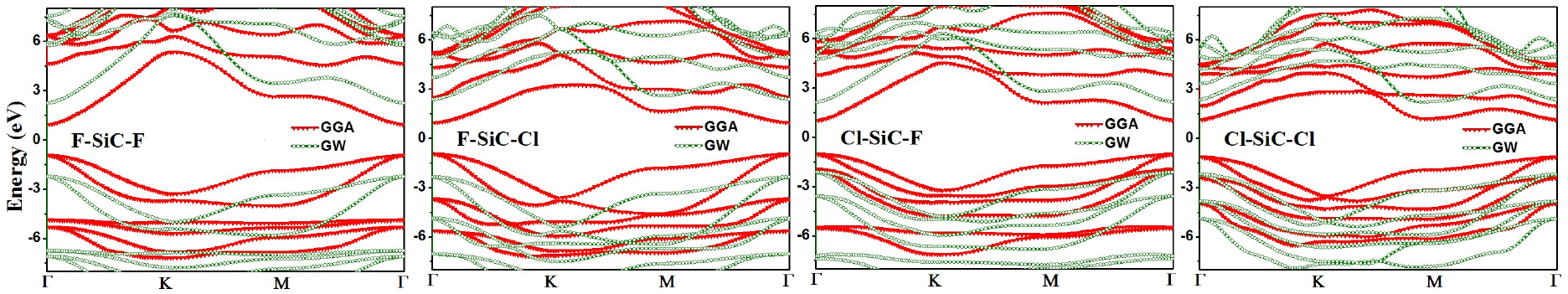}
\end{center}
\par
\vspace{-0.5cm}
\caption{{\protect \small Band structures in Brillouin zone. Red solid
squares and green circles correspond to GGA and GW respectively. Energy is
given with respect to the Fermi energy.}}
\label{band}
\end{figure}
Electronic band structures displayed in Fig.\ref{band} and their
corresponding gap energies given in Table 1 are calculated employing two
different theoretical approaches. Standard GGA-DFT formalism reveals that
all the structures are semi-conductors with a direct gap located at the $%
\Gamma $ point in the Brillouin zone. The highest gap is found for Cl-SiC-Cl
followed by Cl-SiC-F. The structures F-SiC-Cl and F-SiC-F have smaller gap
energies. After quasi-particle corrections using the GW approximation, which
is known to improve the gap description, the band gap becomes indirect
between $\Gamma $ and M points in Cl-SiC-Cl conformer while it remains
direct at $\Gamma $ in the three other configurations. The values of band
gap obtained from GW are much larger than those obtained from GGA. Moreover,
the increasing sequence of the band gaps is completely inverted in
comparison to the GGA derived values since F-SiC-Cl and F-SiC-F have the
highest gap energies of $4.71$eV and $4.47$eV respectively followed by
Cl-SiC-Cl and finally Cl-SiC-F. 
\begin{figure}[tbph]
\begin{center}
\hspace{0cm} \includegraphics[width=15cm]{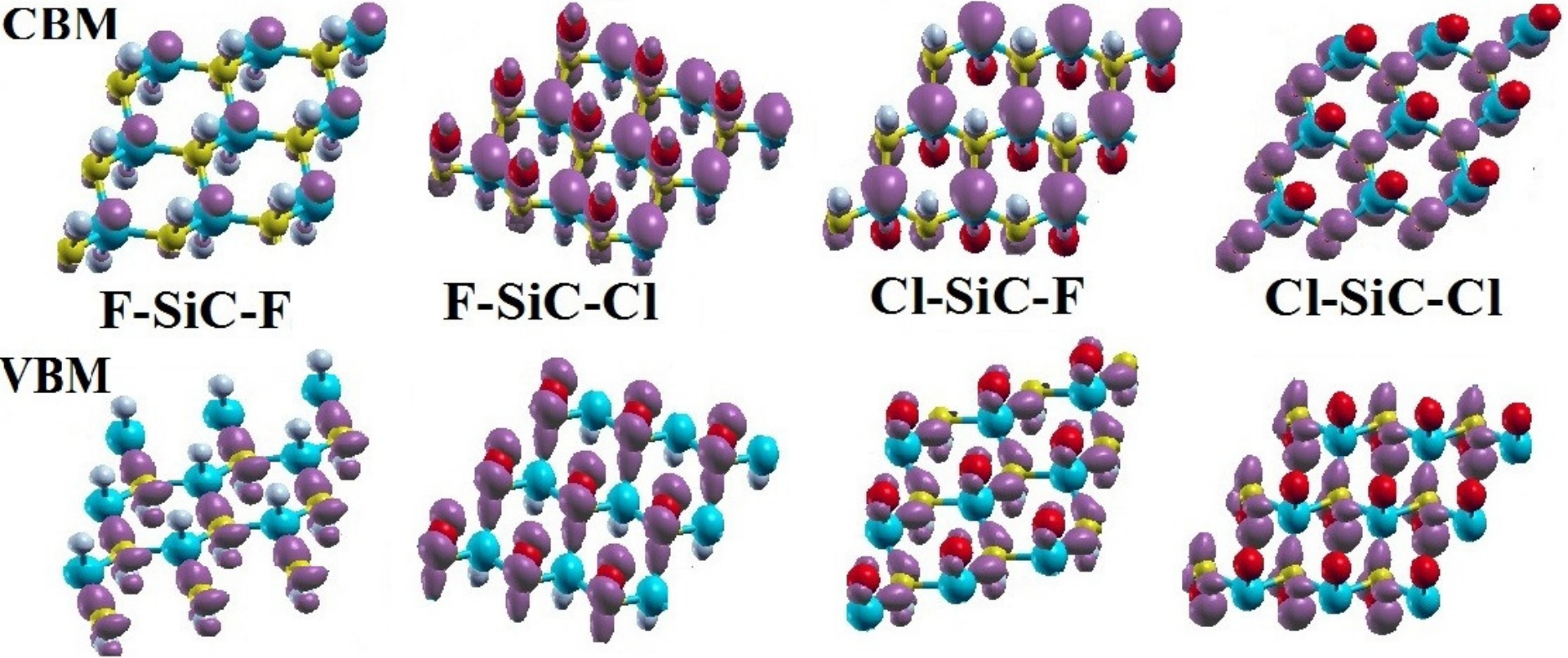}
\end{center}
\par
\vspace{-0.5cm}
\caption{{\protect \small Isosurface charge densities plots describing CBM
(upper) and VBM (lower) corresponding to the four derivatives. The contour
isovalue is set at 40\%. }}
\label{CVbm}
\end{figure}

\begin{figure}[tbph]
\begin{center}
\hspace{0cm} \includegraphics[width=17cm]{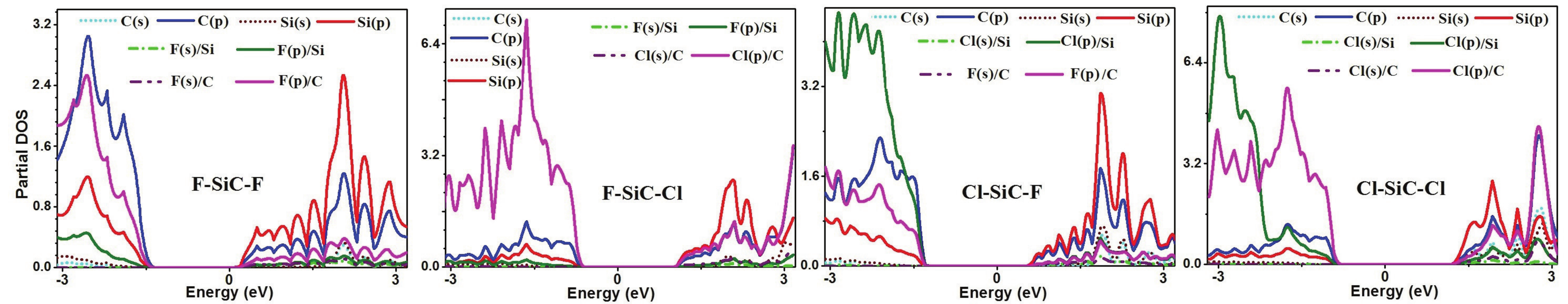}
\end{center}
\par
\vspace{-0.5cm}
\caption{{\protect \small Projected density of states corresponding to s- and
p-orbitals of atoms composing the hybrids. The energy is given with respect
to the Fermi energy.}}
\label{Pdoss}
\end{figure}

To gain further insights into the gaps character, partial density of states
(PDOS) and charge distribution describing the lowest conduction band (CBM)
and the highest valence band (VBM) are plotted. A first analysis of Fig.\ref%
{CVbm} indicates that CBM mainly originates from orbitals of C and Si atoms,
while orbitals of halogen atoms contribute also to VBM. More precisely, the
PDOS of the four structures (see Fig.\ref{Pdoss}) shows that the lower
energy of CBM is dominated by p-electrons of Si-atoms followed by C-atoms.
However, the situation is quite different in VBM. Indeed, in F-SiC-F, VBM is
formed by p-orbitals corresponding to C-atoms, Si-atoms and fluorine that
decorates C-atoms (referred as F$_{/\mathrm{C}}$-atoms). In Cl-SiC-Cl, both
p-orbitals of C-atoms and Cl that bond C-atoms (referred as Cl$_{/\mathrm{C}%
})$ contribute in VBM. In F-SiC-Cl, VBM consists mainly of Cl/C p-orbitals
followed by less involving of C-atoms. Finally, in Cl-SiC-F, VBM is mainly
composed of p-electrons of C- and Cl-atoms that are attached to Si (referred
as Cl$_{/\mathrm{Si}}).$

\begin{figure}[tbph]
\begin{center}
\hspace{0cm} \includegraphics[width=17cm]{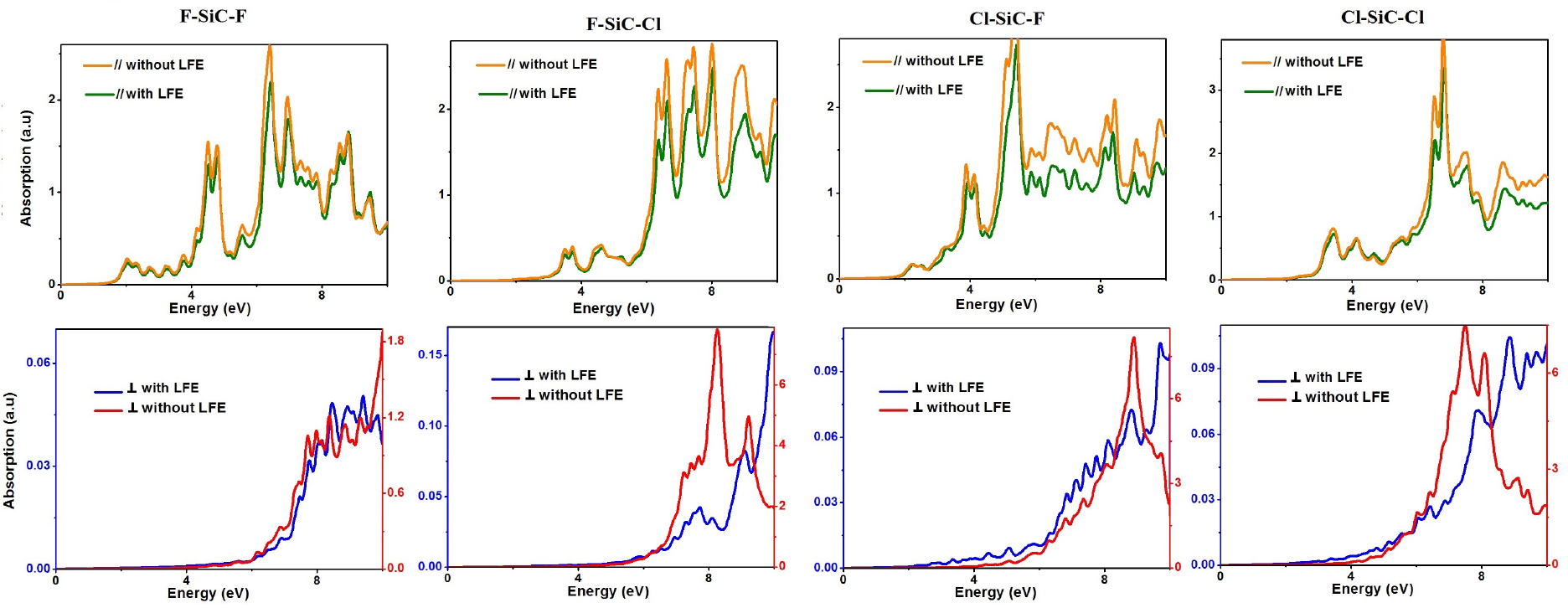}
\end{center}
\par
\vspace{-0.5cm}
\caption{{\protect \small Local field effects (LFE) on optical absorption
spectra for light polarized (top figures) parallel and (down figures)
perpendicular to the hybrids sheet calculated using GGA-RPA.}}
\label{LFE}
\end{figure}

In Fig.\ref{LFE} optical absorption spectra are depicted for light
polarization along and perpendicular to halogenated SiC sheets. Local field
effects (LFE) are also considered using GGA and the Random Phase
Approximation (RPA). Similar to graphene \cite{EE} and pure GeC monolayer 
\cite{nouGeC}, when light polarization is in-plane, LFE have no significant
influence on the species except a very slight reduction in their
intensities. However, effects of local field are more important for
out-of-plane polarization. In this case, LFE shift optical absorption's
curves to higher frequencies and reduce significantly the spectrum intensity
in the four structures. Moreover, in presence of the local field, the
halogenated hybrids F-SiC-F, F-SiC-Cl, Cl-SiC-F and Cl-SiC-Cl are
transparent below $2.04$eV, $2.52$eV, $2.24$eV and $3.44$eV respectively.

\begin{figure}[tbph]
\begin{center}
\hspace{0cm} \includegraphics[width=17cm]{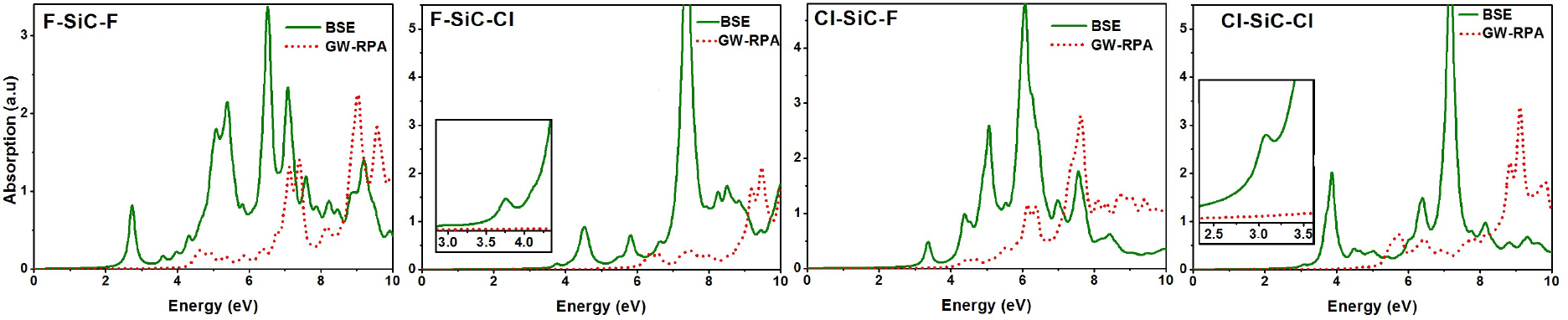}
\end{center}
\par
\vspace{-0.5cm}
\caption{{\protect \small Imaginary part of macroscopic dielectric function
of SiC hybrid calculated using both BSE and GW-RPA. }}
\label{4}
\end{figure}

Optical absorption spectra are determined by the imaginary part of the
macroscopic dielectric function, $Im\left[ \varepsilon (\omega )\right] $.
In the absence of LFE, Fig.\ref{4} displays $Im\left[ \varepsilon (\omega )%
\right] $ calculated either without electron--hole interaction (GW-RPA) or
including excitonic effects from the solution of Bethe--Salpeter equation
(BSE). Only incident light polarized along x--direction is considered. The
effective mass $M_{eff}$ of the bright exciton is given by the expression 
\cite{massex}:%
\begin{equation}
M_{eff}=\frac{E_{b}^{e-h}}{R_{H}}\varepsilon _{r}^{2}m_{0},
\end{equation}

where $R_{H}$ is the Rydberg energy, $\varepsilon _{r}$ is the dielectric
constant, $m_{0}$ is the electron rest mass and $E_{b}^{e-h}$ is the
excitonic binding \ energy. The Bohr radius of this core exciton can also be
calculated with the following equation \cite{massex}%
\begin{equation}
a=\frac{\varepsilon _{r}}{M_{eff}}a_{H}m_{0},
\end{equation}%
where $a_{H}$ is the Bohr radius of hydrogen atom.

The e-h correlations modify dramatically the optical spectra. More
precisely, in the four configurations, excitonic absorption edges are
red-shifted and the spectrum profile is completely different compared to the
GW-RPA spectrum with a main increase in its relative absorption intensity.
Indeed, the main part of the absorption spectra is shifted back to the
GGA-RPA position but with an important redistribution of the resonances due
to correlations and for instance excitonic features, which are inexistent in
RPA.

The first absorption peak is observed at $2.73$eV, $2.96$eV and $3.36$eV for
F-SiC-F, F-SiC-Cl and Cl-SiC-F respectively. It corresponds in the three
cases to optically active (bright) excitonic state. In the three halogenated
structures, the strongly bound excitons result from vertical transitions
between the top of valence band (degenerated) to the bottom of conduction
band at $\Gamma $-point. Thus, they are double degenerate. Their exciton
binding energies, (defined as the difference between the energy of optical
excitation and electronic gap), their Bohr radius and their effective mass
are listed in Table 1. Notice that the exciton binding energy, corresponding
to F-SiC-F, intermediates between fluorographene with $1.96$eV \cite{hh} and
fluorosilicene having $1.48$eV \cite{fluorosilicene}.\newline
The situation is different for full chlorinated SiC that is a semiconductor
with indirect band gap. The first exciton observed in the optical spectrum
is a dark exciton located at $3.04$eV. This strongly bound excitonic state
has an e-h binding energy of $1.34$eV and is due to the reversion of the
oscillator strength of the first active and the first inactive excitons when
the light polarization changes \cite{graphane}. Furthermore, the first
optically active (bright) exciton emerges at $3.07$eV which allowed vertical
transitions from the top two valence bands to the bottom of conduction band
at $\Gamma $-point. This bright exciton is characterized by a binding energy
of $1.31$eV and an effective mass of $0.26m_{0}.$\newline
It follows full fluorination and full chlorination increase optical binding
energy of pristine SiC \cite{drissi2}.

\begin{figure}[tbph]
\begin{center}
\hspace{0cm} \includegraphics[width=14cm]{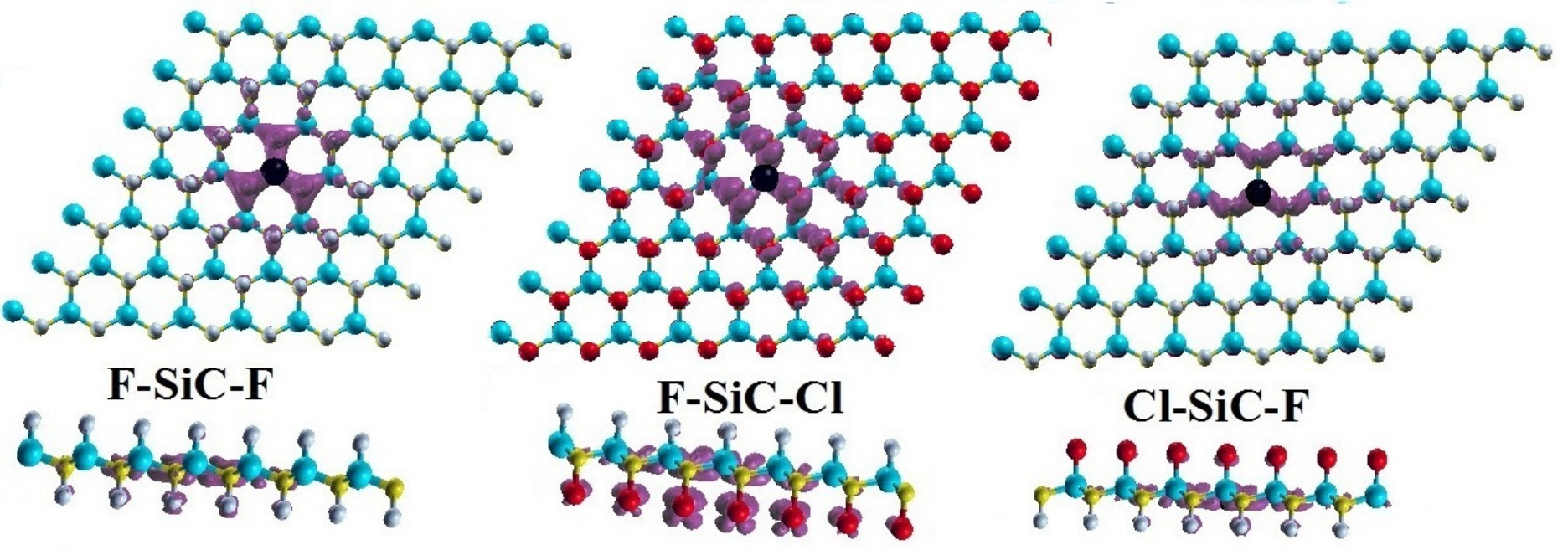}
\end{center}
\par
\vspace{-0.5cm}
\caption{{\protect \small Top and side view of electron probability
distribution with hole position (black circle) fixed above a C atom in
F-SiC-F, F-SiC-Cl and Cl-SiC-F configurations. }}
\label{5}
\end{figure}

\begin{figure}[tbph]
\begin{center}
\hspace{0cm} \includegraphics[width=12cm]{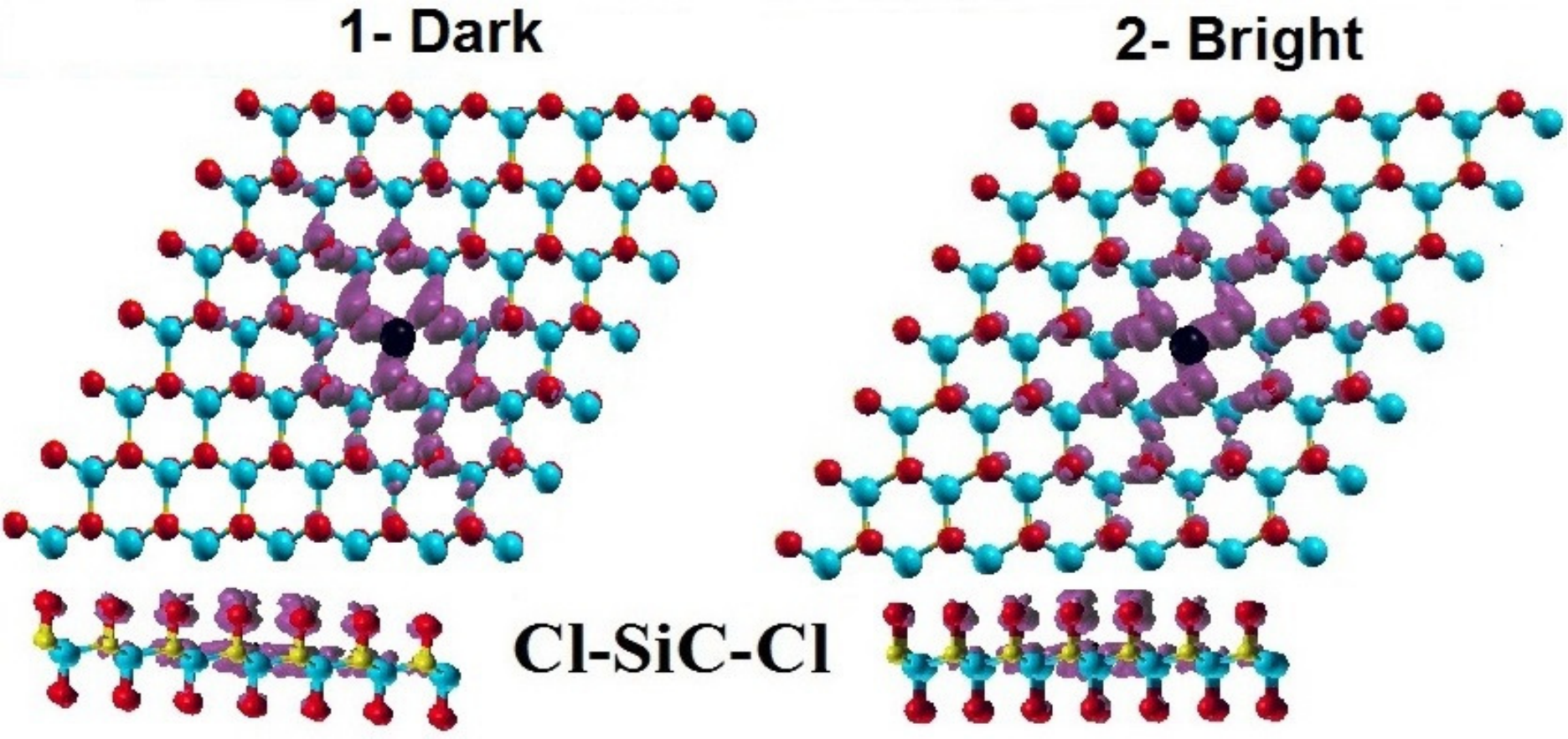}
\end{center}
\par
\vspace{-0.5cm}
\caption{{\protect \small Top and side view of 2D projections of electron
probability distribution in Cl-SiC-Cl structure corresponding to 1- dark
exciton and 2- bright exciton where the hole (black circle) is located above
a C-atom}}
\label{7}
\end{figure}

Figs.\ref{5} and \ref{7} plot electron probability distribution $\left \vert
\psi \left( r_{e},r_{h}\right) \right \vert ^{2}$ to understand correlations
between excited quasi-electron and quasi-hole states in real space. The
coordinate $r_{e}$ refers to electron position and $r_{h}$ is the position
of a hole placed slightly above a Si-atom. The electron-hole amplitude $\Psi
(r_{e};r_{h})$ is invariant to lattice vector shifts when applied
simultaneously to $r_{e}$ and $r_{h}.$

In F-SiC-F and F-SiC-Cl, electron charge density is more localized around
the hole with small radius of $2.34\mathring{A}$ and $2.24\mathring{A}$
respectively compared to $3.36\mathring{A}$ and $4.40\mathring{A}$ obtained
in Cl-SiC-Cl and Cl-SiC-F where the exciton is delocalized along the
material. So, halogenation increases significantly the radius of the exciton
radius of pure SiC. As shown in side view in Figs.\ref{5} and \ref{7},
charge transfer occurs mainly between C- atoms and their adsorbates in the
four structures. This result does not depend on the difference between
electronegativities of the atoms as reported in \cite{hh} and \cite{graphane}%
. It follows that structures where carbon is decorated with Cl, namely
F-SiC-Cl and Cl-SiC-Cl, are promising candidates for possible excitonic
Bose-Einstein condensation as proposed recently in the graphane case \cite%
{graphane}.

Fig. \ref{7} shows a damping behavior for the first exciton in Cl-SiC-Cl,
that is of dark-type. An identical damping nature of distribution is also
observed for the bright exciton. This result reveals a strong excitonic
effect. Moreover, the electron distribution associated to the first bright
exciton is less localized and has a bigger radius compared to the one of
fluorinated-SiC and of F-SiC-Cl. This behavior of the electron probability
distribution of first active and inactive excitons is mainly due to their
small binding energy of $1.31$eV with respect to $1.74$eV and $1.75$eV
obtained for F-SiC-F and F-SiC-Cl respectively. As consequence, the spatial
separation of electron and hole in chlorinated SiC is the largest among the
investigated materials.

\begin{figure}[tbph]
\begin{center}
\hspace{0cm} \includegraphics[width=16cm]{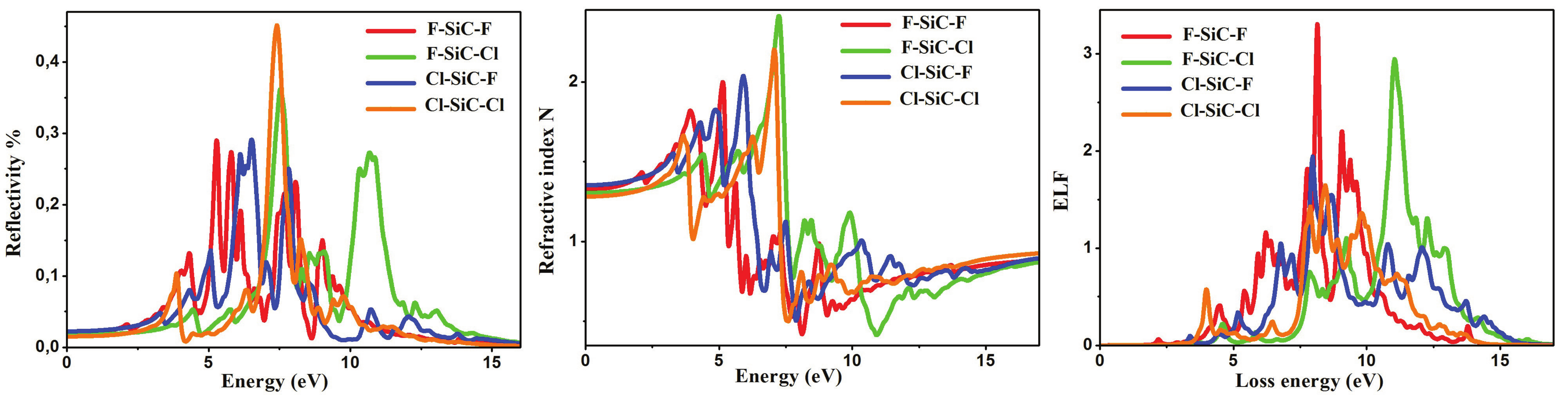}
\end{center}
\par
\vspace{-0.5cm}
\caption{{\protect \small (left) Reflectivity, (center) refraction index and
(right) energy loss function including excitonic effects of halogenated SiC.}
}
\label{AA}
\end{figure}

Other optical quantities such as reflectivity, refractive index and
electron-energy-loss function are derived from the macroscopic dielectric
function $\varepsilon \left( \omega \right) $ that is calculated using GW-BSE%
$.$

At normal incidence, Fresnel reflectivity $R(\omega )$ given by:%
\begin{equation}
R\left( \omega \right) =\left \vert \frac{\sqrt{\varepsilon \left( \omega
\right) }-1}{\sqrt{\varepsilon \left( \omega \right) }+1}\right \vert ^{2}
\end{equation}%
is plotted in Fig.\ref{AA} (left) for the four configurations. It is evident
that these compounds behave like semiconductors since the $R(\omega )$
values are not approach to the unity towards zero energy. The zero frequency
reflectivity $R(0)$ is $0.019\%$, $0.017\%$, $0.015\%$, and $0.022\%$ for of
F-SiC-F, F-SiC-Cl, Cl-SiC-Cl, and Cl-SiC-F respectively. According to the\
reflectivity spectra these materials can be transparent in the visible
region in contrast to the ultraviolet region where more reflectivity occurs.

Fig.\ref{AA} (center) displays the real part of the refractive index $N$
given by $N=n_{+}+in_{-}$ where $n_{+}$ and $n_{-}$ are refractive and
extinction indexes calculated as follows:

\begin{equation}
n_{\pm }\left( \omega \right) =\sqrt{\frac{1}{2}\left( \sqrt{\Re \varepsilon
\left( \omega \right) ^{2}+\Im \varepsilon \left( \omega \right) ^{2}}\pm
\Re \varepsilon \left( \omega \right) \right) }.
\end{equation}%
It is found that the static refraction index at zero energy takes the value
of $1.32$, $1.30$, $1.28$ and $1.35$ for F-SiC-F, F-SiC-Cl, Cl-SiC-Cl, and
Cl-SiC-F respectively. These values are smaller than $1.48$ calculated for
silicon-doped graphene \cite{sic2}. In all configurations, the corresponding
refraction index is minimum where absorption is maximum. Moreover, it is
obvious that the obtained values are inversely proportional to the energy of
direct band gap.

Finally, Fig.\ref{AA} (right) presents energy-loss spectrum $L(\omega ),$
another important optical characteristic that describes the energy loss of a
fast electron crossing the material. It is shown that the plasmon peak
occurs at 8.20 eV, 11.26eV, 8.44eV, and 7.97 eV for F-SiC-F, F-SiC-Cl,
Cl-SiC-Cl, and Cl-SiC-F respectively which corresponds to a rapid decrease
of reflectance in agreement with Fig.\ref{AA} (left). Moreover, the
collective excitation is the point of transition from the metallic to
dielectric property as each material exhibits dielectric behavior above the
plasmon frequency in contrast to the metallic behavior below the plasmon
frequency. Notice also that the direct band gap correlates with the plasmon
frequency.

\section{Conclusion}

In summary, we have studied structural, electronic and optical properties of
four halogenated SiC conformers, namely F-SiC-F, Cl-SiC-Cl, F-SiC-Cl, and
Cl-SiC-F. Phonon dispersion and binding energies reveal that all the
structures are stable which imply their possible fabrication and realization
in laboratory. Whereas GGA-DFT calculations show that halogenation causes
gap energy of SiC to decrease with respect to the pristine case, GW
calculations give larger band gaps for SiC halides. The band gap is indirect
in Cl-SiC-Cl and direct at $\Gamma $ in the three other configurations. The
resulting absorption spectra demonstrate substantial redshifts and
enhancement of absorption peaks compared to the calculated spectra
neglecting excitonic effects. Except Cl-SiC-F that exhibits an exciton
binding energy rather similar to that of pristine SiC, F-SiC-F, Cl-SiC-Cl
and F-SiC-Cl structures have huge binding energy. The strong excitonic
effect in halogenated SiC makes these materials desirable for
opto-electronics applications. Moreover, the charge transfer from carbon to
chlorine in F-SiC-Cl and full chlorinated SiC suggest them as promising
candidates for the Bose-Einstein condensation~\cite{graphane}. The direct
controllable band gaps and the high binding energies make these materials
suitable for optoelctronic applications such as, solar cell, LED, and
batteries while their lower refractive index make them promising for
applications in anti-reflection coatings and high-reflective systems.

\section*{Acknowledgements}

L. B. Drissi and F. Z. Ramadan would like to acknowledge financial support
from the Centre National pour la Recherche Scientifique et Technique
(CNRST)-Morocco. L. B. Drissi and S. Lounis thank the Arab-German Young
academy of Sciences and Humanities (AGYA) and the BMBF for partial funding.

\end{document}